\documentclass[prd,a4paper,showpacs,preprintnumbers]{revtex4} 

\usepackage{epsfig}

\def\nn{\nonumber}        
\def\eqcm{\: ,}           
\def\eqpt{\: .}
\def\adj{{\phantom{h}}}   
\def\bra#1{\mbox{$\langle #1\vert $}}
\def\ket#1{\mbox{$\vert #1\rangle$}}

\newcommand{\bm}[1]{\mbox{\boldmath $#1$}}

\newcommand{\ovl}{\overline}
\newcommand{\open}{{<\kern -0.3 em{\scriptscriptstyle )}}}
\newcommand{\sT}{{\scriptscriptstyle T}}
\newcommand{\nslash}{\kern 0.2 em n\kern -0.45em /}
\newcommand{\Pslash}{\kern 0.2 em P\kern -0.56em \raisebox{0.3ex}{/}}
\newcommand{\pslash}{\kern 0.2 em p\kern -0.4em /}
\newcommand{\kslash}{\kern 0.2 em k\kern -0.45em /}
\newcommand{\Sslash}{\kern 0.2 em S\kern -0.56em \raisebox{0.3ex}{/}}

\newcommand{\ds}{\displaystyle}
\newcommand{\eq}{\begin{equation}}
\newcommand{\ee}{\end{equation}}
\newcommand{\eqa}{\begin{eqnarray}}
\newcommand{\eea}{\end{eqnarray}}

\newcommand{\sumint}{\kern 0.2 em {\textstyle\sum} \kern -1.1 em \int}

\newcommand{\g}{\gamma}
\newcommand{\sig}{\sigma}
\newcommand{\eps}{\epsilon}

\begin{document} 

\title{Interference fragmentation functions in electron-positron annihilation}

\author{Daniel Boer}
\email{dboer@nat.vu.nl}
\affiliation{Dept.\ of Physics and Astronomy, Vrije Universiteit Amsterdam, \\
  De Boelelaan 1081, 1081 HV Amsterdam, The Netherlands}

\author{Rainer Jakob}
\email{rainer.jakob@physik.uni-wuppertal.de}
\affiliation{Fachbereich Physik, Universit\"at Wuppertal, 
  D-42097 Wuppertal, Germany}

\author{Marco Radici}
\email{marco.radici@pv.infn.it}
\affiliation{Dipartimento di Fisica Nucleare e Teorica, 
  Universit\`{a} di Pavia, and\\
  Istituto Nazionale di Fisica Nucleare, 
  Sezione di Pavia, I-27100 Pavia, Italy}

\date{\today}

\begin{abstract}
We study the process of electron-positron annihilation into back-to-back jets,
where in each jet a pair of hadrons is detected. The orientation of these two
pairs with respect to each other can be used to extract the interference
fragmentation functions in a clean way, for instance from BELLE or BABAR
experiment data. This is of relevance for studies of the transversity
distribution function. 

In particular, we focus on two azimuthal asymmetries. The first one has already
been studied by Artru and Collins, but is now expressed in terms of interference 
fragmentation functions. The second asymmetry is new
and involves a function that is related to longitudinal jet handedness.
This asymmetry offers a different way of studying handedness correlations. 
\end{abstract}


\maketitle

\section{\label{sec:intro} Introduction}

Interference fragmentation functions (IFFs) have been suggested as a means to
access transversity~\cite{Ralston:1979ys} 
via single spin asymmetries in $e\, p^\uparrow$ and $p \,
p^\uparrow$ processes, in which the proton is transversely
polarized. Transversity is a measure of how much of the transverse
polarization of a proton is transferred to its quarks. It is a helicity-flip
(or chiral-odd) distribution function that is very hard to measure and thus
far no extraction from data is available. To become sensitive to the
transverse spin of quarks inside a transversely polarized proton one can 
follow two main routes. 
The first one is to use another transversely 
polarized hadron (in initial or final state), like in the Drell-Yan 
process $p^\uparrow\,p^\uparrow \to \ell \, \bar \ell \, X$ or in 
polarized $\Lambda$ hyperon production. 
The second route is to
exploit the possibility that the direction of the transverse polarization
of a fragmenting quark may somehow be encoded in the distribution of hadrons 
inside the resulting jet. 
For instance, the Collins effect~\cite{Collins:1992kk} 
describes the case where the
distribution of a hadron inside the jet follows 
a ${\bm k}_T \times {\bm s}_T$ behaviour, where ${\bm k}_T$ is the 
transverse momentum of the quark
compared to the hadron and ${\bm s}_T$ is the transverse polarization of the
fragmenting quark. Here, transverse means orthogonal to the quark (or,
equivalently, jet) 
direction. Due to the transverse momentum dependence, the Collins 
effect is a very challenging observable both theoretically and experimentally,
and an alternative is formed by the IFFs which 
describe the distribution of two hadrons inside the jet. The idea
is that the orientation of two hadrons with respect to each other and to the
jet direction, is an indicator for the transverse spin direction of the 
quark. Such a correlation is expected to occur due to the strong final state
interactions between the two hadrons: different partial waves can interfere
and this is expected to give rise to nonvanishing, nontrivial fragmentation
functions, the two-hadron IFFs. 

As with all proposals to measure transversity, a second unknown quantity is
introduced, which needs to be measured separately. For the nontrivial 
fragmentation
functions, such as the Collins function and the two-hadron IFFs, the 
cleanest extraction
is from two-jet events in the electron-positron annihilation process. Here, we
will present the leading twist, fully differential cross section for the 
process $e^+ e^- \rightarrow (h_1 h_2)
({\bar h}_1 {\bar h}_2) X$ in terms of products of two-hadron fragmentation
functions. 

In Ref.~\cite{Bianconi:1999cd}, the leading twist, transverse momentum
dependent two-hadron IFFs have been listed. There are two chiral-odd and
one chiral-even IFF, but upon integration over the {\em quark\/} transverse 
momentum dependence only one chiral-odd IFF survives (called $H_1^{\open}$), 
discussed at several places in the 
literature~\cite{Collins:1993kq,Ji:1993vw,Jaffe:1997hf,noiDIS}. 
The relation among the various approaches and to the $\rho$
fragmentation functions (for the two hadrons being two pions) is extensively 
discussed in Ref.~\cite{Bacchetta:2002ux}, 
where the two-hadron final system is
expanded in relative partial waves and a new contribution involving the
transversity at leading twist is identified.

Here, we will discuss the consequences of all three leading twist 
IFFs occurring in 
the process $e^+ e^- \rightarrow (h_1 h_2) ({\bar h}_1 {\bar h}_2) X$; we 
find that upon integrating the differential 
cross section over the {\em observed\/} 
transverse momentum one is actually not only
left with the transverse momentum integrated chiral-odd IFF $H_1^{\open}$, 
but there is also an asymmetry that is governed by the chiral-even IFF,
integrated, but weighted, over the transverse momentum:
\eq
G_1^{\perp}(z,M_h^2) \equiv \int d\xi \int d\phi_R \int d\bm{k}_\sT^{}
\;  \bm{k}_\sT^{}\!\cdot\! {\bm{R}}^{}_\sT \; G_1^{\perp}
(z,\xi,\bm{k}_\sT^2,\bm{R}_\sT^2,\bm{k}^{}_\sT\cdot\bm{R}^{}_\sT),
\ee
where ${\bm{R}}^{}_\sT$ is the transverse part of the 
relative momentum between the two hadrons and $\bm{k}_\sT^{}$ is the quark
transverse momentum (see Sec.~\ref{sec:kin} for the explicit definitions of 
the above quantities).
This function is related (but not identical) to longitudinal jet handedness 
and its resulting asymmetry will be discussed in detail below 
(see Sec.~\ref{sec:asym2}). 

The asymmetry
involving the transverse momentum integrated chiral-odd IFF $H_1^{\open}$
has already been studied in a different (less 
common) notation in a paper by Artru and Collins~\cite{Artru:1995zu}.
It is the asymmetry of present-day 
experimental interest regarding transversity. The extraction of 
$H_1^{\open}$ from the process $e^+ e^- \rightarrow (h_1 h_2) 
({\bar h}_1 {\bar h}_2) X$ is the goal of a group~\cite{GrossePerdekamp}
that will analyze the off-resonance data from the BELLE experiment at KEK. In the 
present article, we provide for the procedure of integrating and properly weighting 
the fully differential cross section to single out the relevant asymmetry. 
The extracted IFF will be of use to several ongoing or starting experiments
aiming to measure transversity in the processes 
$e\, p^\uparrow \to (h_1 h_2) X$ (HERMES, COMPASS) and 
$p\, p^\uparrow \to (h_1 h_2) X$ (RHIC~\cite{GrossePerdekamp}). 

However, the asymmetry involving $G_1^\perp$ also seems of experimental 
interest. It can be viewed as the
chiral-even counterpart of the Artru-Collins asymmetry. An analogous 
asymmetry involving chiral-even fragmentation functions 
does not emerge when only one hadron is detected in each jet; this asymmetry
is thus particular to the multi-hadron fragmentation case. 
But it can also be viewed as an asymmetry arising from a correlation between 
longitudinal jet handedness functions. As such it is relevant 
for single spin asymmetries with longitudinally polarized protons, 
$e\, \vec{p} \to (h_1 h_2) X$ and 
$p\, \vec{p} \to (h_1 h_2) X$, which are proportional to the well-known 
quark helicity distribution function $g_1$ (cf., e.g., Eq.~(31) of 
Ref.~\cite{Bianconi:1999cd}). Since $g_1$ is known to considerable accuracy,
one can extract $G_1^\perp$ from $e\, \vec{p} \to (h_1 h_2) X$ and
actually predict our longitudinal jet handedness correlation in 
$e^+ e^- \rightarrow (h_1 h_2) 
({\bar h}_1 {\bar h}_2) X$, i.e.\ the expression given below in 
Eq.\ (\ref{eq:newasym3}). Any experimental deviation
may be related to a CP-violating effect of the QCD
vacuum~\cite{Efremov:1995ff}.   

The function $G_1^\perp$ is also relevant for the 
studies of IFFs in the processes $e\, p^\uparrow \to (h_1 h_2) X$ and 
$p\, p^\uparrow \to (h_1 h_2) X$. There, next to the asymmetry proportional 
to the transversity function, another $G_1^\perp$ dependent
asymmetry~\cite{noiDIS} occurs, which is  
proportional to the transverse momentum dependent distribution function 
$g_{1T}$~\cite{Tangerman:1994eh}. This function (extrapolated to $x=0$)
gives information on violations of the Burkhardt-Cottingham sum rule.  
Apart from the intrinsic interest in such an asymmetry, it also shows the need
for appropriate weight functions to separate the asymmetry proportional to 
$g_{1T} G_1^\perp$ from the asymmetries proportional to 
$h_1 H_1^\open$ and $h_1 H_1^\perp$ (where $h_1$ denotes the transversity
function).  

The other results presented below, i.e.\ the other terms arising in the 
fully differential $e^+ e^-$ cross section, may also be of
interest in the future and the notation used here hopefully will  
facilitate the communication between different 
experimental groups planning or performing two-hadron IFF-related 
studies for different processes. 

The paper is organized as follows. In Sec.~\ref{sec:kin} we first discuss the
kinematics of the process $e^+ e^- \rightarrow (h_1 h_2) ({\bar h}_1 {\bar
h}_2) X$. In Sec.~\ref{sec:cross} we present the cross section in terms of the
interference fragmentation functions. Next, we investigate extensively the
Artru-Collins azimuthal asymmetry (Sec.~\ref{sec:asym1}) and 
the newly-found longitudinal jet 
handedness asymmetry (Sec.~\ref{sec:asym2}). 
During the discussion of these two asymmetries in 
$e^+ e^- \rightarrow (h_1 h_2) ({\bar h}_1 {\bar h}_2) X$ we also remark on
corresponding asymmetries in two-hadron inclusive 
DIS involving the same IFFs to facilitate comparison.
We end with conclusions (Sec.~\ref{sec:end}).  
 
\begin{figure}[h]
\epsfig{file=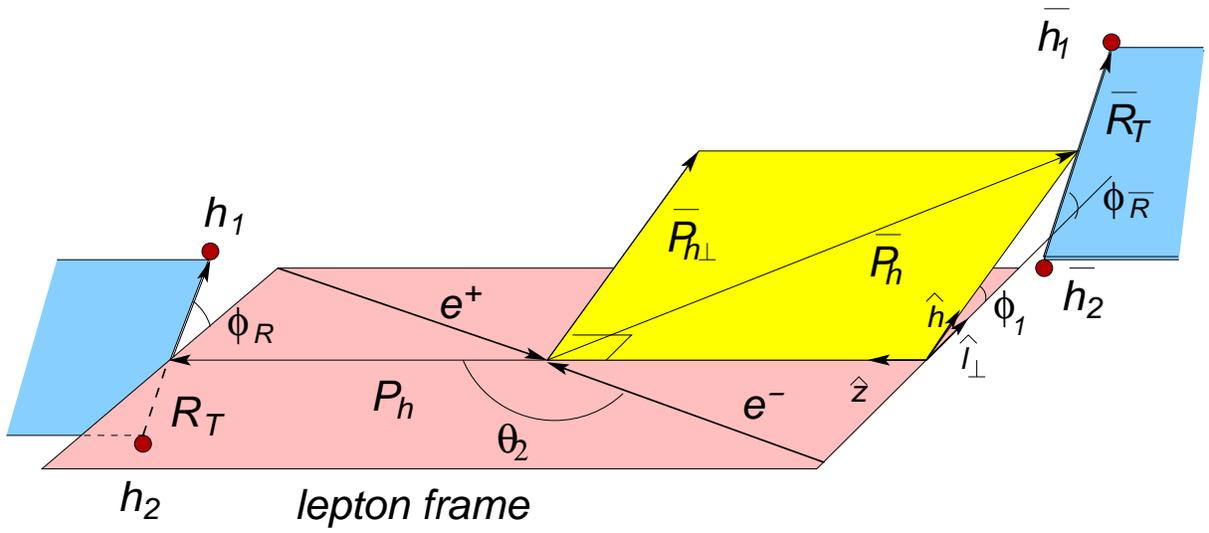,width=16truecm,height=7truecm}
\caption{Kinematic for the 
$e^+ e^- \rightarrow (h_1 h_2) ({\bar h}_1 {\bar h}_2) X$ process.}
\label{fig:kin}
\end{figure}

\section{\label{sec:kin} Kinematics}

We will consider the process 
$e^+e^- \rightarrow (h_1 h_2) ({\bar h}_1 {\bar h}_2) X$,
schematically depicted in Fig.~\ref{fig:kin}. An electron and a positron with
momenta $l$ and $l'$, respectively, annihilate into a photon with timelike
momentum $q=l+l'$ and $q^2=Q^2$. A quark and an antiquark are then emitted and
fragment each one into a residual jet and a pair of leading unpolarized
hadrons $(h_1,h_2)$ with momenta $P_1,P_2$, and masses $M_1,M_2$ (for the 
antiquark we have the corresponding notation  
$({\bar h}_1,{\bar h}_2)$ with momenta ${\ovl P}_1, {\ovl P}_2$ 
and masses ${\ovl M}_1,{\ovl M}_2$). We introduce the vectors 
$P_h=P_1+P_2, \, R=(P_1-P_2)/2$, and 
${\ovl P}_h={\ovl P}_1+{\ovl P}_2, \, {\ovl R}=({\ovl P}_1- {\ovl P}_2)/2$. 
The two jets are emitted in opposite directions, therefore 
$P_h\cdot {\ovl P}_h \sim Q^2$. 
Neglecting $1/Q$ mass-term corrections~\cite{Amst98}, we can 
parametrize the momenta as
\eqa
P_h^\mu &= &\frac{z_h Q}{\sqrt{2}} \  n_-^\mu \nn \\
{\ovl P}_h^\mu &= &\frac{{\ovl z}_h Q}{\sqrt{2}} \   n_+^\mu \nn \\
q^\mu &= &\frac{Q}{\sqrt{2}} \  n_-^\mu 
          + \frac{Q}{\sqrt{2}} \  n_+^\mu + q_T^\mu 
\eqcm
\label{eq:lightpar}
\eea
where $-q_T^2 \equiv Q_T^2 \ll Q^2$, and $n_+, n_-$, are light-like vectors 
satisfying $n_+^2 = n_-^2 = 0$ and $n_+ \cdot n_- = 1$. We use the following
notation, $a^\pm = a \cdot n_\mp$, for a generic 4-vector $a$ with light-cone
components $a=[a^-,a^+,\bm{a}_T]$. We define also  
$z = P_h^-/q^- \sim 2 P_h \cdot q/Q^2 = z_h \, , \, {\ovl z} = 
{\ovl P}_h^+ /q^+ \sim 2 {\ovl P}_h \cdot q/Q^2 = {\ovl z}_h$, representing 
the light-cone momentum fractions of the fragmenting (anti-)quark 
carried by each hadron system. Analogously, 
we define the following fractions 
\eqa
\xi &= &\frac{1}{2}+\frac{R^-}{P_h^-} = \frac{P_1^-}{P_1^-+P_2^-} \nn \\ 
{\ovl \xi} &= &\frac{1}{2}+\frac{{\ovl R}^-}{{\ovl P}_h^-} = 
\frac{{\ovl P}_1^-}{{\ovl P}_1^-+{\ovl P}_2^-}  \eqcm
\label{eq:xi}
\eea
that describe how the momentum of the (anti-)quark is split into each
component of the hadron pair. The $\hat z$ axis is defined using $P_h$; in
particular, from Fig.~\ref{fig:kin} it is $\bm{P}_h \parallel \bm{\hat z}$. 
It is useful to define the so-called $\perp$ plane \cite{Amst97}
perpendicular to $\bm{\hat
z}$, where $\bm{P}_{h\perp} = \bm{q}_\perp = 0$. Up to corrections in 
$Q_T^2/Q^2 \ll 1$, we have ${\ovl P}_{h\, \perp}^\mu = - {\ovl z} q_T^\mu$ 
and, consequently, $\bm{\ovl P}_{h\, \perp} = -{\ovl z} \bm{q}_T$. 
The $\perp$ plane is spanned by the two unit vectors 
\eqa
\hat h &= &\frac{\bm{\ovl P}_{h\, \perp}}{|\bm{\ovl P}_{h\, \perp}|} = 
-\frac{\bm{q}_T}{|\bm{q}_T|} = (\cos \phi_1 \  , \  \sin \phi_1) \nn \\[2mm]
{\hat g}^i &= &\eps_T^{ij} {\hat h}^j \equiv \eps^{-+ij} {\hat h}^j = 
\eps^{0ij3} {\hat h}^j = (\sin \phi_1 \  , \  -\cos \phi_1) \eqcm
\label{eq:versors}
\eea 
with $\phi_1$ defined in Fig.~\ref{fig:kin}. Therefore, we have 
$\hat g \cdot \bm{a} = (\bm{a} \times {\hat h})_z$ for a generic 3-vector
$\bm{a}$. As in Deep Inelastic Scattering (DIS), the 
$\perp$ and transverse $(T)$ components of a 4-vector can be obtained by 
the following tensors 
\eqa
g_T^{\mu \nu} &= &g^{\mu \nu} - n_+^\mu n_-^\nu - n_-^\mu n_+^\nu \nn \\
g_\perp^{\mu \nu} &= &g_T^{\mu \nu} -
 \frac{\sqrt{2} ( n_+^\mu q_T^\nu + n_+^\nu q_T^\mu )}{Q} \eqpt
\eea
In the following, we will consistently neglect the ${\cal O}(1/Q)$ 
difference, thus not distinguishing between $\perp$ and $T$ components of
4-vectors. From previous definitions we have also 
\eqa
|\bm{R}_T|^2 &= &\xi (1-\xi) M_h^2 - (1-\xi) M_1^2 -\xi M_2^2 \nn \\[2mm]
|\bm{\ovl R}_T|^2 &= &{\ovl \xi} (1-{\ovl \xi}) {\ovl M}_h^2 - 
(1-{\ovl \xi}) {\ovl M}_1^2 - {\ovl \xi} {\ovl M}_2^2 \eqcm
\label{eq:rt}
\eea
where $P_h^2=M_h^2$ and ${\ovl P}_h^2={\ovl M}_h^2$ are the invariant masses 
of the two final hadronic systems. The azimuthal angles 
$\phi_R, \phi_{\ovl R}$, parametrizing the transverse components of 
$R,{\ovl R}$, are depicted in Fig.~\ref{fig:kin}. There, a
further azimuthal angle $\phi^l$ should be considered which identifies the 
position of the lepton frame with respect to the laboratory frame. In
Fig.~\ref{fig:kin} it is taken $\phi^l=0$ for convenience, but in the 
following (see Secs.~\ref{sec:asym1} and \ref{sec:asym2}) we will
have to retain its dependence explicitly such that the position 
of the hadron pairs with respect to the lepton frame is $\phi_R - \phi^l$ and 
$\phi_{\ovl R} - \phi^l$, respectively.

The cross section for the 4-unpolarized-particle inclusive $e^+e^-$ annihilation 
is 
\eq
\frac{2P_1^0 \, 2P_2^0 \, 2{\ovl P}_1^0 \, 2{\ovl P}_2^0 \, d\sig }
     {d\bm{P}_1 d\bm{P}_2 d\bm{\ovl P}_1 d\bm{\ovl P}_2} =  
     \frac{\alpha^2}{Q^6} \, L_{\mu \nu} W^{\mu \nu}_{(4h)} \eqcm
\label{eq:x}
\ee
where 
\eqa
L^{\mu \nu} &=& Q^2 \Big[ -2\,A(y)\, g_\perp^{\mu \nu} 
      + 4\, B(y) \, {\hat z}^\mu {\hat z}^\nu \, 
      - 4\,B(y) \left( {\hat l}_\perp^\mu {\hat l}_\perp^\nu \,
      + \frac{1}{2}\, g_\perp^{\mu \nu} \right)  \nn \\
& &\quad {}- \, 2 \, C(y) B^{1/2}(y) ({\hat z}^\mu
{\hat l}_\perp^\nu + {\hat z}^\nu {\hat l}_\perp^\mu ) \Big] 
\eqcm \label{eq:Ltens} 
\\[2mm]
A(y) & = & \left(\frac{1}{2}-y+y^2\right) \ \stackrel{cm}{=} \; 
\ds{\frac{\left( 1 + \cos^2\theta_2 \right)}{4}}  \nn\\[2mm]
B(y) & = & y\,(1-y) \ \stackrel{cm}{=} \ \ds{\frac{\sin^2 \theta_2}{4}}
\nn\\[2 mm]
C(y) & = & 1-2y \ \stackrel{cm}{=} \ - \cos \theta_2 
\eqcm \label{eq:Ltenscoff}
\eea
is the lepton tensor. In fact, only the $L^{\mu \nu}_{\perp}$ 
part contributes at
leading twist. The invariant $y = P_h\cdot l/P_h\cdot q\sim l^-/q^-$
becomes, in the lepton center-of-mass (cm) frame, $y=(1+\cos \theta_2)/2$,
where $\theta_2$ is defined in Fig.~\ref{fig:kin}. The unit vectors are
defined as  
\eqa
{\hat l}_\perp^\mu &= &l^\mu_\perp / |\bm{l}_\perp | \nn \\
{\hat z}^\mu &= & \frac{2}{z Q} P_h - \frac{q^\mu}{Q} \eqpt
\label{eq:versors2}
\eea

The hadronic tensor is
\eqa
W^{\mu \nu}_{(4h)}(q;P_1,P_2,{\ovl P}_1,{\ovl P}_2) 
&= &\frac{1}{(2\pi)^{10}} \, 
\sumint_X \frac{d\bm{P}_X}{(2\pi )^32P_X^0} \, (2\pi )^4 \delta(q-P_X-P_1-P_2-
{\ovl P}_1 - {\ovl P}_2) \nn \\[1mm] 
& &\mbox{\hspace{1.5truecm}} \times \bra{0}J^\mu(0)
\ket{P_X;P_1,P_2,{\ovl P}_1,{\ovl P}_2} 
\bra{P_X;P_1,P_2,{\ovl P}_1,{\ovl P}_2} J^\nu(0) \ket{0} \eqpt
\label{eq:Wtens4}
\eea
Such definition allows for recovering the corresponding formulae in the case 
of 2-particle, 1-particle and totally inclusive $e^+e^-$ annihilation. For
example, after integrating over one of the two hadrons in each pair, 
\eqa
\sumint_{h_2} \frac{d\bm{P}_2}{2P_2^0} \, \sumint_{{\bar h}_2} 
\frac{d\bm{\ovl P}_2}{2{\ovl P}_2^0} \, 
\frac{2P_1^0 \, 2P_2^0 \, 2{\ovl P}_1^0 \, 2{\ovl P}_2^0 d\sig}
{d\bm{P}_1 d\bm{P}_2 d\bm{\ovl P}_1 d\bm{\ovl P}_2} &\equiv & 
\frac{2P_1^0 \, 2{\ovl P}_1^0 d\sig}{d\bm{P}_1 d\bm{\ovl P}_1} = 
\frac{\alpha^2}{Q^6}\, L_{\mu \nu} \, \sumint_{h_2} \frac{d\bm{P}_2}{2P_2^0} 
\sumint_{{\bar h}_2} \frac{d\bm{\ovl P}_2}{2{\ovl P}_2^0}\, 
W^{\mu\nu}_{(4h)} \nn \\
& &\mbox{\hspace{-3truecm}} 
=\frac{\alpha^2}{Q^6}\, L_{\mu \nu} \, \frac{1}{(2\pi )^4} 
\, \sumint_X \frac{d\bm{P}_X}{(2\pi )^3 2P_X^0} \, (2\pi )^4 
\delta(q-P_X-P_1-P_2-{\ovl P}_1-{\ovl P}_2) \nn \\
& &\mbox{\hspace{-2truecm}} \times 
\sumint_{h_2} \frac{d\bm{P}_2}{(2\pi )^3 2P_2^0} 
\sumint_{{\bar h}_2} \frac{d\bm{\ovl P}_2}{(2\pi )^3 2{\ovl P}_2^0} \, 
\bra{0} J^\mu \ket{X,P_1,P_2,{\ovl P}_1,{\ovl P}_2} 
\bra{X,P_1,P_2,{\ovl P}_1,{\ovl P}_2} J^\nu \ket{0} \nn \\
& &\mbox{\hspace{-3truecm}} =\frac{\alpha^2}{Q^6}\, L_{\mu \nu} \, 
\frac{1}{(2\pi )^4} 
\, \sumint_{X'} \frac{d\bm{P}_{X'}}{(2\pi )^3 2P_{X'}^0} \, (2\pi )^4 
\delta(q-P_{X'}-P_1-{\ovl P}_1) \, 
\sumint_{h_2} \frac{d\bm{P}_2}{(2\pi )^3 2P_2^0} 
\nn \\
& &\mbox{\hspace{-2.6truecm}} \times \sumint_{{\bar h}_2} 
\frac{d\bm{\ovl P}_2}{(2\pi )^3 2{\ovl P}_2^0}\, \bra{0} J^\mu 
\ket{X'-P_2-{\ovl P}_2,P_1,P_2,{\ovl P}_1,{\ovl P}_2} 
\bra{X'-P_2-{\ovl P}_2,P_1,P_2,{\ovl P}_1,{\ovl P}_2} J^\nu \ket{0} \nn \\
& &\mbox{\hspace{-3truecm}} 
=\frac{\alpha^2}{Q^6}\, L_{\mu \nu} \, \frac{1}{(2\pi )^4} 
\, \sumint_{X'} \frac{d\bm{P}_{X'}}{(2\pi )^3 2P_{X'}^0} \, (2\pi )^4 
\delta(q-P_{X'}-P_1-{\ovl P}_1) \nn \\
& &\times \bra{0} J^\mu \ket{X',P_1,{\ovl P}_1} 
\bra{X',P_1,{\ovl P}_1} J^\nu \ket{0} \nn \\
& &\mbox{\hspace{-3truecm}} 
\equiv \frac{\alpha^2}{Q^6}\, L_{\mu \nu} \, W^{\mu\nu}_{(2h)} 
\eqcm
\label{eq:checke+e-4-2}
\eea
we recover the cross section for the 2-particle inclusive $e^+e^-$
annihilation~\cite{Amst97} after the replacement 
${\ovl P}_1 \leftrightarrow P_2$. Further integrations over the detected 
hadrons lead to the 1-particle inclusive and totally inclusive cross 
sections (cf.~\cite{Amst97}).

Consistently with Eq.~(\ref{eq:lightpar}), we have
\eqa
P_h^+ \ll P_h^- &\rightarrow &P_h^0 \sim \frac{1}{\sqrt{2}} P_h^- \nn \\
R^+ \ll R^- &\rightarrow &R^0 \sim \frac{1}{\sqrt{2}} R^- \nn \\
\frac{R^-}{P_h^-} &= &\xi - \frac{1}{2} \sim \frac{R^0}{P_h^0} \equiv
\frac{E_R}{E_h} \nn \\
{\ovl P}_h^- \ll {\ovl P}_h^+ &\rightarrow &{\ovl P}_h^0 \sim 
\frac{1}{\sqrt{2}} {\ovl P}_h^+ \nn \\
{\ovl R}^- \ll {\ovl R}^+ &\rightarrow &{\ovl R}^0 \sim 
\frac{1}{\sqrt{2}} {\ovl R}^+ \nn \\
\frac{{\ovl R}^+}{{\ovl P}_h^+} &\equiv &{\ovl \xi} - \frac{1}{2} \sim 
\frac{{\ovl R}^0}{{\ovl P}_h^0} \equiv \frac{{\ovl E}_R}{{\ovl E}_h} 
\eqpt \label{eq:lightappr}
\eea
The elementary phase space can then be transformed as follows:
\eqa
\frac{d\bm{P}_1 d\bm{P}_2 d\bm{\ovl P}_1 d\bm{\ovl P}_2}
{2P_1^0 \, 2P_2^0 \, 2{\ovl P}_1^0 \, 2{\ovl P}_2^0} &= &
\frac{d\bm{P}_h \, d\bm{R} \, d\bm{\ovl P}_h \, d\bm{\ovl R}}
{(E_h^2-4E_R^2)({\ovl E}_h^2-4{\ovl E}_R^2)} \sim  
\frac{|\bm{P}_h|^2 d|\bm{P}_h| d\Omega_h \, 
d\bm{R}_T dR^- \, d\bm{\ovl P}_{h\perp} 
d{\ovl P}_h^+ d\bm{\ovl R}_T d{\ovl R}^+}
{8\sqrt{2} \, (P_h^-)^2 \xi (1- \xi) 
({\ovl P}_h^+)^2 {\ovl \xi} (1- {\ovl \xi})} \nn \\
&\sim & \frac{z Q^2}{16 \, \xi (1- \xi)}\,dz\,d\Omega_h\,d\bm{R}_T\,d\xi\, 
\frac{1}{4 {\ovl z} {\ovl \xi} (1- {\ovl \xi})} 
d\bm{\ovl P}_{h\perp}\,d{\ovl z}\,d\bm{\ovl R}_T\,d{\ovl \xi} \eqcm
\label{eq:phasesp}
\eea
where $d\Omega_h = 2 dy d\phi^l$, since $\bm{P}_{h\perp} = 0$ and its
azimuthal angle actually defines the position of the lepton plane with respect
to the laboratory frame.  
Using Eq.~(\ref{eq:rt}) and 
\eqa
d\bm{R}_T &= &J \, d\phi_R \, dM_h^2 \quad \mbox{with} \, J=\left\vert 
\begin{array}{cc} 
   \displaystyle{\frac{\partial R_{Tx}}{\partial M_h^2} = 
                 \frac{\xi (1-\xi)}{2|\bm{R}_T|}} \cos\phi_R  & \; 
   \displaystyle{\frac{\partial R_{Tx}}{\partial \phi_R}} = 
                 -|\bm{R}_T| \sin\phi_R 
   \\[3mm]
  \displaystyle{\frac{\partial R_{Ty}}{\partial M_h^2} = 
                \frac{\xi (1-\xi)}{2|\bm{R}_T|} \sin\phi_R } & \; 
  \displaystyle{\frac{\partial R_{Ty}}{\partial \phi_R}} = 
                |\bm{R}_T| \cos\phi_R
  \end{array} \right\vert = \frac{\xi (1-\xi)}{2} \nn \\
d\bm{\ovl R}_T &= & \frac{{\ovl \xi} (1-{\ovl \xi})}{2} \, 
d\phi_{\ovl R} \, d{\ovl M}_h^2 \eqcm
\label{eq:drt}
\eea
the cross section can be rewritten as
\eq
\frac{d\sig}{d\bm{q}_T \, dz \, d\xi \, d\phi_R \, dM_h^2 \, d{\ovl z} \, 
d{\ovl \xi} \, d\phi_{\ovl R} \, d{\ovl M}_h^2 \, dy \, d\phi^l} =  
\frac{\alpha^2}{128 \, Q^4} \, z {\ovl z} \, L_{\mu \nu} 
W^{\mu\nu}_{(4h)} \eqpt
\label{eq:x2}
\ee


\section{\label{sec:cross} Cross section}

\subsection{\label{sec:w} The hadronic tensor}

To leading order the expression for the hadron tensor is
\eq
W^{\mu\nu}_{(4h)} 
\sim 3 \, (32)^2 \, z {\ovl z} \sum_{a,\ovl a} e_a^2 \, \int 
d\bm{k}_\sT^{} d\bm{\ovl k}_\sT^{} \, \delta^2(\bm{k}_\sT^{}+
\bm{\ovl k}_\sT^{}-\bm{q}_\sT^{})\, \mbox{Tr} \left[ \left( 
\frac{1}{32{\ovl z}} \int d{\ovl k}^{\,-} {\ovl \Delta} 
\right)_{{\ovl k}_\adj^+={\ovl P}_h^+/{\ovl z}} \, \g^\mu \, \left( 
\frac{1}{32z} \int dk^+ \Delta \right)_{k_\adj^-=P_h^-/z} \, \g^\nu \right] 
\eqpt
\label{eq:Wmunu}
\ee
The (partly integrated) correlation function $\Delta$ is parametrized in terms
of fragmentation functions as \cite{noiDIS}
\eqa 
\lefteqn{ 
  \left. \frac{1}{32z}\int dk^+ \Delta(k;P_h,R) 
                         \right|_{k_\adj^-=P_h^-/z, {\bf k}_\sT} } 
  \nn\\[5pt]
&&
= \frac{1}{4\pi} \, \frac{1}{4} \left\{ 
 D_1^a (z,\xi,\bm{k}_\sT^2,\bm{R}_\sT^2,\bm{k}_\sT\cdot\bm{R}_\sT)\, 
 \nslash_- 
 {}- G_1^{\perp \, a} 
     (z,\xi,\bm{k}_\sT^2,\bm{R}_\sT^2,\bm{k}_\sT\cdot\bm{R}_\sT)\, 
 \frac{\eps_{\mu\nu\rho\sig}\,\g^\mu\,n_-^\nu\,k_\sT^\rho\,R_\sT^\sig}
            {M_1 M_2}\,
 \g_5  \right. \nn\\[5pt]
&&
\left. {}+ H_1^{\open \, a} 
           (z,\xi,\bm{k}_\sT^2,\bm{R}_\sT^2,\bm{k}_\sT\cdot\bm{R}_\sT)
 \, \frac{\sig_{\mu\nu}\, R_\sT^\mu\, n_-^\nu}{M_1+M_2} \, {}+ \, 
 H_1^{\perp \, a} 
 (z,\xi,\bm{k}_\sT^2,\bm{R}_\sT^2,\bm{k}_\sT\cdot\bm{R}_\sT)\, 
 \frac{\sig_{\mu\nu}\, k_\sT^\mu\, n_-^\nu}{M_1+M_2} \right\} \eqpt
 \label{eq:delta}
\eea
We parametrize the antiquark correlation function $\ovl \Delta$ in the same way
by employing overlined quantities, but in this case the 
suppressed ${\ovl k}{}^{\,-}$ component is integrated over.

At leading twist, we have the usual nice probabilistic interpretation of the
fragmentation functions in Eq.~(\ref{eq:delta}): $D_1^a$ is the probability 
for an unpolarized quark with flavor $a$ to fragment into the unpolarized
hadron pair ($h_1,h_2$), $G_1^{\perp\, a}$ is the probability difference for a
longitudinally polarized quark with flavor $a$ and opposite chiralities to
fragment into ($h_1,h_2$), both $H_1^{\open\, a}, \, H_1^{\perp\, a}$, give
the same probability difference but for a transversely polarized fragmenting
quark. $G_1^{\perp\, a}, \, H_1^{\open\, a}, \, H_1^{\perp\, a}$, are all {\it
naive} T-odd and $H_1^{\open\, a}, \, H_1^{\perp\, a}$ are furthermore chiral
odd. The function $H_1^{\perp\, a}$ represents a generalization of the Collins
effect, namely for two hadrons instead of one. However, $H_1^{\open\, a}$
originates from a genuinely new effect, because it relates the transverse
polarization of the fragmenting quark to the angular distribution of the
hadron pair in the $\perp$ plane 
(defined in Sec.~\ref{sec:kin}~\cite{noiDIS}).

\vspace{.8truecm}


\begin{figure}[h]
\epsfig{file=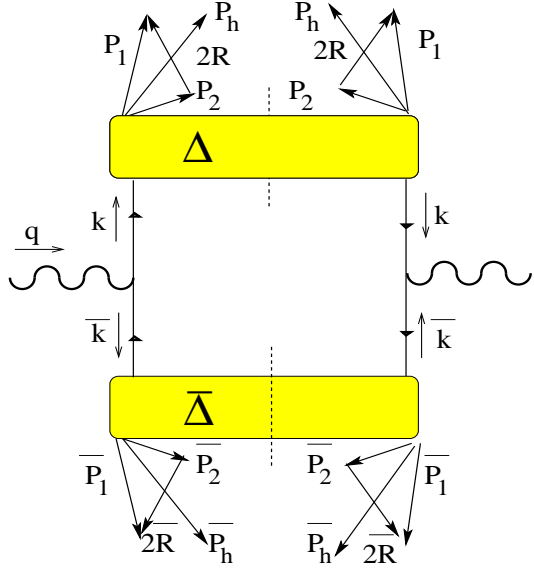,width=7truecm,height=7.5truecm}
\caption{Leading-twist contribution to 4p-inclusive $e^+e^-$ annihilation.}
\label{fig:handbag}
\end{figure}


\subsection{\label{sec:dsig} The fully differential cross section}

For the case of the $e^+e^-$ annihilation into four unpolarized (or spinless) hadrons
with two leading hadrons in each current jet (see Fig.~\ref{fig:handbag} for a
diagrammatic representation at leading order), the differential cross section at 
leading order in $1/Q$ and $\alpha_s$ is
(including now summation over flavor indices with quark charges $e_a$ in
units of $e$)
\eqa
\lefteqn{\frac{d\sigma (e^+e^-\to (h_1h_2)(\bar h_1\bar h_2)X)}
              {d\bm{q}_\sT^{}\,dz\,d\xi\,dM_h^2\,d\phi^{}_R\, d{\ovl z}\,
	       d{\ovl \xi}\, d{\ovl M}_h^2\,d\phi^{}_{\ovl R}\, dy\, 
	       d\phi^l} 
	 = \sum_{a,\ovl a}\; e_a^2 \, \frac{6\,\alpha^2}{Q^2} \;z^2 {\ovl z}^2
	 \;\Bigg\{  A(y) \, {\cal F}\left[D_1^a\ovl D{}_1^a \right]} 
	 \nn\\[5pt]
&&
{}+ \cos(2\phi_1)\;B(y)\;{\cal F}\left[\left(2\,\hat{\bm{h}}\!\cdot\!\bm k_\sT^{}\,
   \,\hat{\bm{h}}\!\cdot\!\bm{\ovl k}_\sT^{}\,-\,\bm k_\sT^{}\!\cdot \!
   \bm{\ovl k}_\sT^{}\,\right)
   \frac{H_1^{\perp a} {\ovl H}_1^{\perp a}}
        {(M_1+M_2)({\ovl M}_1+{\ovl M}_2)} \right] \nn\\[5pt] 
&& {}- \sin(2\phi_1)\;B(y)\;{\cal F}\left[\left(\,\hat{\bm{h}}\!\cdot\!\bm k_\sT^{}\,
   \,\hat{\bm{g}}\!\cdot\!\bm{\ovl k}_\sT^{}\,+\,\hat{\bm{h}}\!\cdot\!
   \bm{\ovl k}_\sT^{}\, \,\hat{\bm{g}}\!\cdot\!\bm k_\sT^{}\,\right)
   \frac{H_1^{\perp a} {\ovl H}_1^{\perp a}}
        {(M_1+M_2)({\ovl M}_1+{\ovl M}_2)} \right] \nn\\[5pt] 
&& {}+ \cos(\phi_R+\phi_{\ovl R}-2\phi^l)\;B(y)\;|\bm{R}_\sT|\,
   |\bm{\ovl R}_\sT|\;{\cal F}\left[
    \frac{H_1^{\open a} {\ovl H}_1^{\open a}}
         {(M_1+M_2)({\ovl M}_1+{\ovl M}_2)} \right] \nn\\[5pt] 
&&
{}+\cos(\phi_1 + \phi_R-\phi^l)\;B(y)\;|\bm{R}_\sT|\; {\cal F}\left[\,\bm{\hat h}\!
   \cdot \!\bm{\ovl k}_\sT^{}\,\frac{H_1^{\open a} {\ovl H}_1^{\perp a}}
   {(M_1+M_2)({\ovl M}_1+{\ovl M}_2)} \right] \nn\\[5pt]
&&
{}-\sin(\phi_1 + \phi_R-\phi^l)\;B(y)\;|\bm{R}_\sT|\; {\cal F}\left[\,\bm{\hat g}\!
   \cdot \!\bm{\ovl k}_\sT^{}\,\frac{H_1^{\open a} {\ovl H}_1^{\perp a}}
   {(M_1+M_2)({\ovl M}_1+{\ovl M}_2)} \right] \nn\\[5pt]
&&
{}+ \cos(\phi_1 + \phi_{\ovl R}-\phi^l)\;B(y)\;|\bm{\ovl R}_\sT|\; 
   {\cal F}\left[\,\bm{\hat h}\!\cdot \!\bm{k}_\sT^{}\,
   \frac{H_1^{\perp a} {\ovl H}_1^{\open a}}
        {(M_1+M_2)({\ovl M}_1+{\ovl M}_2)} \right] \nn\\[5pt]
&&
{}- \sin(\phi_1 + \phi_{\ovl R}-\phi^l)\;B(y)\;|\bm{\ovl R}_\sT|\; 
   {\cal F}\left[\,\bm{\hat g}\!\cdot \!\bm{k}_\sT^{}\,
   \frac{H_1^{\perp a} {\ovl H}_1^{\open a}}
        {(M_1+M_2)({\ovl M}_1+{\ovl M}_2)} \right] \nn\\[5pt]
&&
{}+A(y)\;|\bm{R}_\sT|\,|\bm{\ovl R}_\sT|\;\Bigg[ 
    \sin(\phi_1-\phi_R+\phi^l)\,\sin(\phi_1-\phi_{\ovl R}+\phi^l)\;{\cal F}\left[
    \,\bm{\hat h}\!\cdot \!\bm{k}_\sT^{}\,\bm{\hat h}\!\cdot \!\bm{\ovl k}_\sT^{}\,
    \frac{G_1^{\perp a} {\ovl G}_1^{\perp a}}
         {M_1M_2 {\ovl M}_1 {\ovl M}_2} \right] \nn\\[5pt]
&&
{}+\sin(\phi_1-\phi_R+\phi^l)\,\cos(\phi_1-\phi_{\ovl R}+\phi^l)\;{\cal F}\left[
    \,\bm{\hat h}\!\cdot \!\bm{k}_\sT^{}\,\bm{\hat g}\!\cdot \!\bm{\ovl k}_\sT^{}\,
    \frac{G_1^{\perp a} {\ovl G}_1^{\perp a}}
         {M_1M_2 {\ovl M}_1 {\ovl M}_2} \right] \nn\\[5pt]
&&
{}+\cos(\phi_1-\phi_R+\phi^l)\,\sin(\phi_1-\phi_{\ovl R}+\phi^l)\;{\cal F}\left[
    \,\bm{\hat g}\!\cdot \!\bm{k}_\sT^{}\,\bm{\hat h}\!\cdot \!\bm{\ovl k}_\sT^{}\,
    \frac{G_1^{\perp a} {\ovl G}_1^{\perp a}}
         {M_1M_2{\ovl M}_1{\ovl M}_2} \right] \nn\\[5pt]
&&
{}+\cos(\phi_1-\phi_R+\phi^l)\,\cos(\phi_1-\phi_{\ovl R}+\phi^l)\;{\cal F}\left[
    \,\bm{\hat g}\!\cdot \!\bm{k}_\sT^{}\,\bm{\hat g}\!\cdot \!\bm{\ovl k}_\sT^{}\,
    \frac{G_1^{\perp a}{\ovl G}_1^{\perp a}}
         {M_1M_2{\ovl M}_1{\ovl M}_2} \right]  \Bigg] \Bigg\} \eqcm
\label{eq:fulldx}
\eea
where the convolution $\cal F$ is defined as
\eqa
{\cal F}\left[ w(\bm k_\sT^{},\bm{\ovl k}_\sT^{})\;D^a {\ovl D}^a \, \right] 
&\equiv &\int d\bm k_\sT^{}\; d\bm{\ovl k}_\sT^{}\; \delta^2 
(\bm{\ovl k}_\sT^{}+ \bm k_\sT^{}-\bm q_\sT^{}) \; w(\bm k_\sT^{},
\bm{\ovl k}_\sT^{})\; 
D^a(z,\xi,\bm{k}_\sT^2,\bm{R}_\sT^2,\bm{k}_\sT^{}\cdot\bm{R}_\sT^{}) \nn \\[5pt]
& &\mbox{\hspace{6.5truecm}} 
{\ovl D}^a ({\ovl z},{\ovl \xi},\bm{\ovl k}_\sT^2,
\bm{\ovl R}_\sT^2, \bm{\ovl k}_\sT^{}\cdot\bm{\ovl R}_\sT^{}) \eqpt
\label{eq:convol}
\eea

The azimuthal dependence is dictated by the fact that any information about
the 
azimuthal asymmetry of the distribution of the four hadrons must be encoded by
the 
relative position of $\bm{R}_T$ and $\bm{\ovl R}_T$ with respect to the 
lepton frame, i.e.\ by the $\phi_R-\phi^l$ and $\phi_{\ovl R}-\phi^l$ angles, 
respectively, and by the azimuthal position of the lepton frame 
itself.


\section{\label{sec:asym1} The Artru-Collins azimuthal asymmetry}

In this Section, we will obtain an azimuthal asymmetry in the distribution of the 
four hadrons that arises only due to the transverse relative momenta of each pair, 
i.e.\ only due to the relative position of each pair plane with respect to the 
lepton plane (see Fig.~\ref{fig:kin}). For this purpose, the cross section of 
Eq.~(\ref{eq:fulldx}) must be properly weighted and its dependence on the intrinsic 
transverse momenta of
the quarks integrated out. We present the procedure in considerable detail, 
since this will form a crucial aspect of a practical analysis of 
experimental data. We will show that only $H_1^{\open}$ survives the 
integration, which is the same fragmentation function appearing in the single-spin 
asymmetry that can be built at leading twist in the case of two-hadron inclusive 
DIS~\cite{Ji:1993vw,Jaffe:1997hf,noiDIS}. Therefore, under the hypothesis of 
factorization (collinear factorization in this particular case), the combined 
analysis of the two semi-inclusive processes allows in principle 
to deduce the fragmentation function from $e^+e^-$ and then disentangle the 
transversity distribution in the corresponding DIS cross section at leading twist.

We define the asymmetry
\eqa
A(y,z,{\ovl z},M_h^2, {\ovl M}_h^2) &= & 
   \frac{\langle\cos(\phi_R^{}+\phi_{\ovl R}^{}-2\phi^l) \rangle}
        {\langle 1 \rangle } \nn \\[3mm]
&\equiv &
\Bigg[ \int d\xi \int d{\ovl \xi} \int_0^{2\pi} \frac{d\phi^l}{2\pi} \int_0^{2\pi} 
d\phi_R^{} \int_0^{2\pi} d\phi_{\ovl R}^{} \, 
\cos(\phi_R^{} + \phi_{\ovl R}^{}-2\phi^l)  \nn \\[2mm]
& &\mbox{\hspace{1truecm}} \times \int d\bm{q}_\sT^{} \, 
     \frac{d\sig (e^+e^-\to (h_1h_2)(\bar h_1\bar h_2)X)}
          {dy\,d\phi^l\,dz\,d{\ovl z}\,d\xi\,d{\ovl \xi}\,d{\bm q_\sT^{}}\,
	   dM_h^2\,d\phi^{}_R\, d{\ovl M}_h^2\,d\phi^{}_{\ovl R}} \Bigg] \nn 
	   \\[2mm]
& &\times \Bigg[ \int d\xi \int d{\ovl \xi} \int_0^{2\pi} \frac{d\phi^l}{2\pi} 
\int_0^{2\pi} d\phi_R^{} \int_0^{2\pi} d\phi_{\ovl R}^{} \nn \\[2mm]
& &\mbox{\hspace{1truecm}} \times \int d\bm{q}_\sT^{} \, 
     \frac{d\sig (e^+e^-\to (h_1h_2)(\bar h_1\bar h_2)X)}
          {dy\,d\phi^l\,dz\,d{\ovl z}\,d\xi\,d{\ovl \xi}\,d{\bm q_\sT^{}}\,
	   dM_h^2\,d\phi^{}_R\, d{\ovl M}_h^2\,d\phi^{}_{\ovl R}} \Bigg]^{-1} 
	   \eqpt
\label{eq:asym}
\eea

Let us consider first the term in the numerator, involving the trigonometric weight. 
If we insert the cross section of Eq.~(\ref{eq:fulldx}) into it and consider just 
the average over the azimuthal positions $\phi^l$ of the lepton plane, we can see 
that the first three terms give a vanishing contribution because they involve the 
integral
\eq
\int_0^{2\pi} \frac{d\phi^l}{2\pi} \, \cos (\phi_R + \phi_{\ovl R} - 2\phi^l) = 0 
\eqpt \label{eq:num1-3}
\ee
The fourth term gives
\eqa
& &\sum_{a,\ovl a}\; e_a2 \, \frac{6\,\alpha^2}{Q^2} \;z^2 {\ovl z}^2\; 
B(y) \int d\xi \int d{\ovl \xi} \int_0^{2\pi} d\phi_R^{} \int_0^{2\pi} 
d\phi_{\ovl R}^{} \int_0^{2\pi} \frac{d\phi^l}{2\pi} \; |\bm{R}^{}_\sT| 
\, |\bm{\ovl R}^{}_\sT| \; \cos^2 (\phi^{}_R+\phi^{}_{\ovl R}-2\phi^l) 
\int d\bm{q}^{}_\sT \nn \\[2mm]
& &\mbox{\hspace{4truecm}} \times {\cal F}\left[
\frac{H_1^{\open a}{\ovl H}_1^{\open a}}
     {(M_1+M_2)({\ovl M}_1+{\ovl M}_2)} \right] = \nn \\[2mm]
& &\sum_{a,\ovl a}\; e_a^2 \, \frac{6\,\alpha^2}{Q^2} \;z^2 {\ovl z}^2\; 
B(y) \int d\xi \int d{\ovl \xi} \int_0^{2\pi} d\phi_R^{} \int_0^{2\pi} 
d\phi_{\ovl R}^{} \; \frac{|\bm{R}^{}_\sT| \, |\bm{\ovl R}^{}_\sT|}
                               {2(M_1+M_2)({\ovl M}_1+{\ovl M}_2)} 
\int d\bm{q}^{}_\sT \nn \\[2mm]
& &\mbox{\hspace{2truecm}} \times \int d\bm k_\sT^{}\; d\bm{\ovl k}_\sT^{}\; 
\delta^2 (\bm{\ovl k}_\sT^{}+\bm k_\sT^{}-\bm q_\sT^{}) \; H_1^{\open \,a} 
(z,\xi,\bm{k}_\sT^2,\bm{R}_\sT^2,\bm{k}^{}_\sT\cdot\bm{R}^{}_\sT) \; 
{\ovl H}_1^{\open \, a} ({\ovl z},{\ovl \xi},\bm{\ovl k}_\sT^2,
\bm{\ovl R}_\sT^2,\bm{\ovl k}_\sT^{}\cdot\bm{\ovl R}_\sT^{}) = \nn \\[2mm]
& &\sum_{a,\ovl a}\; e_a^2 \, \frac{3\alpha^2}{Q^2} \;z^2 {\ovl z}^2\; 
\ds{\frac{B(y)}{(M_1+M_2)({\ovl M}_1+{\ovl M}_2)}} H_{1\,(R)}^{\open \,a} 
(z,M_h^2) \; {\ovl H}_{1\, (R)}^{\open \, a}({\ovl z},{\ovl M}_h^2) \eqcm
\label{eq:num4}
\eea
where 
\eqa
H_{1\, (R)}^{\open \,a} (z,M_h^2) &= &\int d\xi \, |\bm{R}_\sT^{}| \int_0^{2\pi} 
d\phi_R^{} \int d\bm{k}_\sT^{} \; H_1^{\open \,a} (z,\xi,\bm{k}_\sT^2,\bm{R}_\sT^2,
\bm{k}^{}_\sT\cdot\bm{R}^{}_\sT) \nn \\
{\ovl H}_{1\, (R)}^{\open \, a}({\ovl z},{\ovl M}_h^2) &= &\int 
d{\ovl \xi} \; |\bm{\ovl R}^{}_\sT| \int_0^{2\pi} d\phi_{\ovl R}^{} \int 
d\bm{\ovl k}_\sT^{} \; {\ovl H}_1^{\open \, a} ({\ovl z},{\ovl \xi},
\bm{\ovl k}_\sT^2,\bm{\ovl R}_\sT^2,\bm{\ovl k}_\sT^{}\cdot
\bm{\ovl R}_\sT^{}) 
\label{eq:h1angle}
\eea
are the same moments of fragmentation functions that appear in the following 
leading twist single-spin asymmetry arising in two-hadron semi-inclusive DIS 
off a transversely polarized target (see Eq.~(17) of Ref.~\cite{noiDIS}):
\eq
\langle \sin(\phi_R^{} -2\phi^l) \rangle \propto B(y) \, |\bm{S}_\perp | \; 
\sum_a e_a^2\; x \; h_1^a(x) \; H_{1\, (R)}^{\open \, a} (z,M_h^2) \eqcm
\label{eq:h1angledis}
\ee
where $\bm{S}_\perp$ is the transverse polarization of the target and $x$ is 
the light-cone momentum fraction of the quark.

The fifth through eighth 
terms give again a vanishing contribution, because they involve
integrals of the kind
\eq
\int_0^{2\pi} \frac{d\phi^l}{2\pi} \; \left\{ \begin{array}{c} \cos 2\phi^l \\  
\sin 2\phi^l \end{array} \right\} \otimes \left\{ \begin{array}{c} \cos \phi^l \\  
\sin \phi^l \end{array} \right\} \, = \, 0 \eqpt
\label{eq:num5-8}
\ee
Finally, it is instructive to check that the last four terms in Eq.~(\ref{eq:fulldx})
vanish only after the combined effect of the average over $d\phi^l$ and the 
integral upon $d\bm{q}_\sT^{}$. In fact, 
\eqa
\sum_{a,\ovl a}\; e_a^2 \, \frac{6\,\alpha^2}{Q^2} \;z^2 {\ovl z}^2\; &
A(y) &\int d\xi \int d{\ovl \xi} \int_0^{2\pi} d\phi_R^{} 
\int_0^{2\pi} d{\ovl \phi}_R^{} \int_0^{2\pi} \frac{d\phi^l}{2\pi} \; 
\frac{|\bm{R}^{}_\sT| \, |\bm{\ovl R}^{}_\sT|}
     {M_1 M_2 {\ovl M}_1 {\ovl M}_2} \; 
\cos (\phi^{}_R+\phi^{}_{\ovl R}-2\phi^l) \int d\bm{q}^{}_\sT \nn \\[3pt]
& &\times \Bigg\{ \sin(\phi_1-\phi_R+\phi^l)\,\sin(\phi_1-\phi_{\ovl R}+\phi^l)\;
  {\cal F}\left[ \,\bm{\hat h}\!\cdot \!\bm{k}_\sT^{}\,\bm{\hat h}\!\cdot \!
   \bm{\ovl k}_\sT^{}\, G_1^{\perp a}{\ovl G}_1^{\perp a} \right] + \nn
   \\[2pt]
& &\qquad \sin(\phi_1-\phi_R+\phi^l)\,\cos(\phi_1-\phi_{\ovl R}+\phi^l)\;
  {\cal F}\left[ \,\bm{\hat h}\!\cdot \!\bm{k}_\sT^{}\,\bm{\hat g}\!\cdot \!
   \bm{\ovl k}_\sT^{}\, G_1^{\perp a}{\ovl G}_1^{\perp a} \right] + \nn 
   \\[2pt]
& &\qquad \cos(\phi_1-\phi_R+\phi^l)\,\sin(\phi_1-\phi_{\ovl R}+\phi^l)\;
  {\cal F}\left[ \,\bm{\hat g}\!\cdot \!\bm{k}_\sT^{}\,\bm{\hat h}\!\cdot \!
   \bm{\ovl k}_\sT^{}\, G_1^{\perp a}{\ovl G}_1^{\perp a} \right] + \nn 
   \\[2pt]
& &\qquad \cos(\phi_1-\phi_R+\phi^l)\,\cos(\phi_1-\phi_{\ovl R}+\phi^l)\;
  {\cal F}\left[ \,\bm{\hat g}\!\cdot \!\bm{k}_\sT^{}\,\bm{\hat g}\!\cdot \!
   \bm{\ovl k}_\sT^{}\, G_1^{\perp a}{\ovl G}_1^{\perp a} \right] \Bigg\} = 
   \nn \\[8pt]
\sum_{a,\ovl a}\; e_a^2 \, \frac{6\,\alpha^2}{Q^2} \;z^2 {\ovl z}^2\; &
A(y) &\int d\xi \int d{\ovl \xi} \int_0^{2\pi} d\phi_R^{} \int_0^{2\pi} 
d\phi_{\ovl R}^{} \; \frac{|\bm{R}^{}_\sT| \, |\bm{\ovl R}^{}_\sT|}
                               {M_1 M_2 {\ovl M}_1 {\ovl M}_2} \; 
\int d\bm{q}^{}_\sT  \int d\bm k_\sT^{}\; d\bm{\ovl k}_\sT^{}\; \delta^2 
(\bm{\ovl k}_\sT^{}+\bm k_\sT^{}-\bm q_\sT^{}) \nn \\[5pt]
& &\times \left\{ \frac{1}{4} (\cos^2\phi_1 - \sin^2\phi_1) \; 
   (\bm{\hat g}\!\cdot \!\bm{k}_\sT^{}\,\bm{\hat g}\!\cdot \!\bm{\ovl k}_\sT^{} - 
    \bm{\hat h}\!\cdot \!\bm{k}_\sT^{}\,\bm{\hat h}\!\cdot \!\bm{\ovl k}_\sT^{}) +
    \right. \nn \\[5pt]
& &\qquad \left. \frac{1}{4} \, \sin\phi_1 \, \cos\phi_1 \; 
    (\bm{\hat h}\!\cdot \!\bm{k}_\sT^{}\,\bm{\hat g}\!\cdot \!\bm{\ovl k}_\sT^{} + 
    \bm{\hat g}\!\cdot \!\bm{k}_\sT^{}\,\bm{\hat h}\!\cdot \!\bm{\ovl k}_\sT^{})
    \right\} \; G_1^{\perp a}{\ovl G}_1^{\perp a} = \nn \\[8pt]
\sum_{a,\ovl a}\; e_a^2 \, \frac{3\,\alpha^2}{2 Q^2} \;z^2 {\ovl z}^2\; &
A(y) &\int d\xi \int d{\ovl \xi} \int_0^{2\pi} d\phi_R^{} \int_0^{2\pi} 
d\phi_{\ovl R}^{} \; \frac{|\bm{R}^{}_\sT| \, |\bm{\ovl R}^{}_\sT|}
                               {M_1 M_2 {\ovl M}_1 {\ovl M}_2} 
\int d\bm k_\sT^{}\; d\bm{\ovl k}_\sT^{} \nn \\[5pt]
& &\times \left\{ 
  \frac{(\bm k_\sT^{}+\bm{\ovl k}_\sT^{})_y^2-
        (\bm k_\sT^{}+\bm{\ovl k}_\sT^{})_x^2}
       {|\bm k_\sT^{}+\bm{\ovl k}_\sT^{}|^4} \; 
  \left[ (\bm k_\sT^{}\times \bm{\ovl k}_\sT^{})_z^2 + 
     \bm k_\sT^{}\cdot (\bm k_\sT^{}+\bm{\ovl k}_\sT^{}) \; 
     \bm{\ovl  k}_\sT^{}\cdot (\bm k_\sT^{}+\bm{\ovl k}_\sT^{}) \right] + 
     \right. \nn \\[5pt]
& &\qquad \left. 2 \; 
   \frac{(\bm k_\sT^{}+\bm{\ovl k}_\sT^{})_x^{}\;
         (\bm k_\sT^{}+\bm{\ovl k}_\sT^{})_y^{}}
        {|\bm k_\sT^{}+\bm{\ovl k}_\sT^{}|^4} \; 
	(\bm k_\sT^{}\times \bm{\ovl k}_\sT^{})_z^{} \; 
	(\bm{\ovl k}_\sT^2 - \bm k_\sT^2) \right\} \nn \\[5pt]
& &\qquad \times G_1^{\perp \,a} (z,\xi,\bm{k}_\sT^2,\bm{R}_\sT^2,
    \bm{k}^{}_\sT\cdot\bm{R}^{}_\sT) \; 
    {\ovl G}_1^{\perp \, a} ({\ovl z},{\ovl \xi},\bm{\ovl k}_\sT^2,
    \bm{\ovl R}_\sT^2,\bm{\ovl k}_\sT^{}\cdot\bm{\ovl R}_\sT^{}) = 0 
    \eqcm
\label{eq:num9-12}
\eea
because the last two integrands are odd under the transformations
\eqa
k_{\sT \,x} \leftrightarrow k_{\sT \,y} &\quad , \quad &
{\ovl k}_{\sT \,x} \leftrightarrow {\ovl k}_{\sT \,y} \nn \\[1mm]
R_{\sT \,x} \leftrightarrow R_{\sT \,y} &\quad , \quad &
{\ovl R}_{\sT \,x} \leftrightarrow {\ovl R}_{\sT \,y} \eqpt
\label{eq:transf}
\eea
Hence, we have
\eqa
\langle\cos(\phi_R^{}+\phi_{\ovl R}^{}-2\phi^l) \rangle &= &
\int d\xi \int d{\ovl \xi} \int_0^{2\pi} \frac{d\phi^l}{2\pi} \int_0^{2\pi} 
d\phi_R^{} \int_0^{2\pi} d\phi_{\ovl R}^{} \, 
\cos(\phi_R^{} + \phi_{\ovl R}^{}-2\phi^l) \nn \\[2mm]
& &\mbox{\hspace{1truecm}} \times \int d\bm{q}_\sT^{} \, 
     \frac{d\sig (e^+e^-\to (h_1h_2)(\bar h_1\bar h_2)X)}
          {dy\,d\phi^l\,dz\,d{\ovl z}\,d\xi\,d{\ovl \xi}\,d{\bm q_\sT^{}}\,
	   dM_h^2\,d\phi^{}_R\, d{\ovl M}_h^2\,d\phi^{}_{\ovl R}} \nn \\[2mm]
&= &\sum_{a,\ovl a}\; e_a^2 \, \frac{3\alpha^2}{Q^2} \;
\frac{z^2 {\ovl z}^2 \; B(y)}{(M_1+M_2)({\ovl M}_1+{\ovl M}_2)} \; 
H_{1\, (R)}^{\open \,a} (z,M_h^2) \; {\ovl H}_{1\, (R)}^{\open \, a}
({\ovl z},{\ovl M}_h^2) \eqpt
\label{eq:numtot}
\eea

In a similar way, it is straightforward to prove that the unweighted cross 
section receives a 
contribution only from the first term of Eq.~(\ref{eq:fulldx}), i.e.
\eqa
\langle 1 \rangle &= &
\int d\xi \int d{\ovl \xi} \int_0^{2\pi} \frac{d\phi^l}{2\pi} \int_0^{2\pi} 
d\phi_R^{} \int_0^{2\pi} d\phi_{\ovl R}^{} \nn \\[2mm]
& &\mbox{\hspace{1truecm}} \times \int d\bm{q}_\sT^{} \, 
     \frac{d\sig (e^+e^-\to (h_1h_2)(\bar h_1\bar h_2)X)}
          {dy\,d\phi^l\,dz\,d{\ovl z}\,d\xi\,d{\ovl \xi}\,d{\bm q_\sT^{}}\,
	   dM_h^2\,d\phi^{}_R\, d{\ovl M}_h^2\,d\phi^{}_{\ovl R}} \nn \\[2mm]
&= &\sum_{a,\ovl a}\; e_a^2 \, \frac{6 \alpha^2}{Q^2} \; A(y)\; z^2 \, 
   {\ovl z}^2\; D_1^a (z,M_h^2) \; {\ovl D}_1^a ({\ovl z},
   {\ovl M}_h^2) \eqcm
\label{eq:dentot}
\eea
where
\eqa
D_1^a (z,M_h^2) &= &\int d\xi \int_0^{2\pi} d\phi_R^{} \int d\bm{k}_\sT^{} \; 
D_1^a (z,\xi,\bm{k}_\sT^2,\bm{R}_\sT^2,\bm{k}^{}_\sT\cdot\bm{R}^{}_\sT) \nn \\
{\ovl D}_1^a({\ovl z},{\ovl M}_h^2) &= &\int d{\ovl \xi} 
\int_0^{2\pi} d\phi_{\ovl R}^{} \int d\bm{\ovl k}_\sT^{} \; 
{\ovl D}_1^a ({\ovl z},{\ovl \xi},\bm{\ovl k}_\sT^2,
\bm{\ovl R}_\sT^2,\bm{\ovl k}_\sT^{}\cdot\bm{\ovl R}_\sT^{}) 
\label{eq:d1}
\eea
are the same fragmentation functions as 
arising in the unweighted cross section 
at leading twist for the two-hadron semi-inclusive DIS process
(see Eq.~(18) of Ref.~\cite{noiDIS}). 

The final expression for the azimuthal asymmetry is, from
Eq.~(\ref{eq:asym}),  
\eqa
A(y,z,{\ovl z},M_h^2, {\ovl M}_h^2) &= & \frac{1}{2} \; \left[ 
\sum_{a,\ovl a}\; e_a^2 \;
\frac{z^2 {\ovl z}^2\; B(y)}{(M_1+M_2)({\ovl M}_1+{\ovl M}_2)} \; 
H_{1\, (R)}^{\open \,a} (z,M_h^2) \; {\ovl H}_{1\, (R)}^{\open \, a}
({\ovl z},{\ovl M}_h^2) \right] \nn \\[2mm]
& &\times \left[ \sum_{a,\ovl a}\; e_a^2 \, z^2 \, {\ovl z}^2 \, A(y)\; 
D_1^a (z,M_h^2) \; {\ovl D}_1^a({\ovl z},{\ovl M}_h^2) \right]^{-1} \eqpt
\label{eq:asym2}
\eea
This azimuthal asymmetry is our version of the Artru-Collins 
asymmetry~\cite{Artru:1995zu}.


\section{\label{sec:asym2} The longitudinal jet handedness azimuthal asymmetry}

The other azimuthal asymmetry we will explicitly derive is defined as
\eqa
A(y,z,{\ovl z},M_h^2, {\ovl M}_h^2) &= & 
\frac{\langle\cos(2(\phi_R^{}-\phi_{\ovl R}^{})) \rangle}
     {\langle 1 \rangle} \nn \\[3mm]
&\equiv &
\Bigg[ \int d\xi \int d{\ovl \xi} \int_0^{2\pi} \frac{d\phi^l}{2\pi} \int_0^{2\pi} 
d\phi_R^{} \int_0^{2\pi} d\phi_{\ovl R}^{} \, 
\cos(2(\phi_R^{} - \phi_{\ovl R}^{}))  \nn \\[2mm]
& &\mbox{\hspace{1truecm}} \times \int d\bm{q}_\sT^{} \, 
     \frac{d\sig (e^+e^-\to (h_1h_2)(\bar h_1\bar h_2)X)}
          {dy\,d\phi^l\,dz\,d{\ovl z}\,d\xi\,d{\ovl \xi}\,d{\bm q_\sT^{}}\,
	   dM_h^2\,d\phi^{}_R\, d{\ovl M}_h^2\,d\phi^{}_{\ovl R}} \Bigg] \nn 
	   \\[2mm]
& &\times \Bigg[ \int d\xi \int d{\ovl \xi} \int_0^{2\pi}
\frac{d\phi^l}{2\pi} \int_0^{2\pi} d\phi_R^{} \int_0^{2\pi} d\phi_{\ovl R}^{} 
\nn \\[2mm]
& &\mbox{\hspace{1truecm}} \times \int d\bm{q}_\sT^{} \, 
     \frac{d\sig (e^+e^-\to (h_1h_2)(\bar h_1\bar h_2)X)}
          {dy\,d\phi^l\,dz\,d{\ovl z}\,d\xi\,d{\ovl \xi}\,d{\bm q_\sT^{}}\,
	   dM_h^2\,d\phi^{}_R\, d{\ovl M}_h^2\,d\phi^{}_{\ovl R}} \Bigg]^{-1} 
	   \eqpt
\label{eq:newasym}
\eea
Note that this asymmetry is independent of the orientation of the lepton scattering 
plane, contrary to the asymmetry of the previous Section. 

By performing the integrations in the same order as in
Eqs.~(\ref{eq:num1-3}-\ref{eq:num9-12}), the surviving terms are
\eqa
\sum_{a,\ovl a}\; e_a^2 \, \frac{6\,\alpha^2}{Q^2} \;z^2 {\ovl z}^2 & 
A(y) &\int d\xi \int d{\ovl \xi} \int_0^{2\pi} d\phi_R^{} 
\int_0^{2\pi} d\phi_{\ovl R}^{}\int_0^{2\pi} \frac{d\phi^l}{2\pi} \; 
\frac{|\bm{R}^{}_\sT| \, |\bm{\ovl R}^{}_\sT|}
     {M_1 M_2 {\ovl M}_1 {\ovl M}_2}\; 
\cos(2(\phi_R-\phi_{\ovl R})) \int d\bm{q}^{}_\sT \nn \\[3pt]
& &\times \Bigg\{ \sin(\phi_1-\phi_R+\phi^l)\,\sin(\phi_1-\phi_{\ovl R}+\phi^l)\;
  {\cal F}\left[ \,\bm{\hat h}\!\cdot \!\bm{k}_\sT^{}\,\bm{\hat h}\!\cdot \!
   \bm{\ovl k}_\sT^{}\, G_1^{\perp a}{\ovl G}_1^{\perp a} \right] + \nn
   \\[2pt]
& &\qquad \sin(\phi_1-\phi_R+\phi^l)\,\cos(\phi_1-\phi_{\ovl R}+\phi^l)\;
  {\cal F}\left[ \,\bm{\hat h}\!\cdot \!\bm{k}_\sT^{}\,\bm{\hat g}\!\cdot \!
   \bm{\ovl k}_\sT^{}\, G_1^{\perp a}{\ovl G}_1^{\perp a} \right] + \nn 
  \\[2pt]
& &\qquad \cos(\phi_1-\phi_R+\phi^l)\,\sin(\phi_1-\phi_{\ovl R}+\phi^l)\;
  {\cal F}\left[ \,\bm{\hat g}\!\cdot \!\bm{k}_\sT^{}\,\bm{\hat h}\!\cdot \!
   \bm{\ovl k}_\sT^{}\, G_1^{\perp a}{\ovl G}_1^{\perp a} \right] + \nn 
  \\[2pt]
& &\qquad \cos(\phi_1-\phi_R+\phi^l)\,\cos(\phi_1-\phi_{\ovl R}+\phi^l)\;
  {\cal F}\left[ \,\bm{\hat g}\!\cdot \!\bm{k}_\sT^{}\,\bm{\hat g}\!\cdot \!
   \bm{\ovl k}_\sT^{}\, G_1^{\perp a}{\ovl G}_1^{\perp a} \right] \Bigg\} = 
   \nn \\[8pt]
\sum_{a,\ovl a}\; e_a^2 \, \frac{6\,\alpha^2}{Q^2} \;z^2 {\ovl z}^2\; & 
A(y) &\int d\xi \int d{\ovl \xi} \int_0^{2\pi} d\phi_R^{} 
\int_0^{2\pi} d\phi_{\ovl R}^{} \; 
\frac{1}{2M_1 M_2 {\ovl M}_1 {\ovl M}_2}
\; \cos(2(\phi_R-\phi_{\ovl R})) \nn \\[3pt]
& &\mbox{\hspace{-3.5truecm}} \times \Bigg\{ \cos(\phi_R-\phi_{\ovl R})\;
\bm{\hat R}^{}_\sT \!\cdot\! \bm{\hat {\ovl R}}^{}_\sT \int d\bm{k}_\sT^{}
\;  \bm{k}_\sT^{}\!\cdot\! \bm{R}^{}_\sT \, G_1^{\perp a}
(z,\xi,\bm{k}_\sT^2,\bm{R}_\sT^2,\bm{k}^{}_\sT\cdot\bm{R}^{}_\sT) \; 
\int d\bm{\ovl k}_\sT^{}\; 
\bm{\ovl k}_\sT^{}\!\cdot\! \bm{\ovl R}^{}_\sT {\ovl G}_1^{\perp a}
({\ovl z},{\ovl \xi},\bm{\ovl k}_\sT^2,\bm{\ovl R}_\sT^2,
\bm{\ovl k}^{}_\sT\cdot\bm{\ovl R}^{}_\sT)  \nn \\[2pt]
& &\mbox{\hspace{-3truecm}} + \sin(\phi_R-\phi_{\ovl R})\;
\bm{\hat R}^{}_\sT \!\times\! \bm{\hat {\ovl R}}^{}_\sT \int d\bm{k}_\sT^{}
\; \bm{k}_\sT^{}\!\cdot\! \bm{R}^{}_\sT \, G_1^{\perp a}
(z,\xi,\bm{k}_\sT^2,\bm{R}_\sT^2,\bm{k}^{}_\sT\cdot\bm{R}^{}_\sT) \; 
\int d\bm{\ovl k}_\sT^{} \; 
\bm{\ovl k}_\sT^{}\!\cdot\! \bm{\ovl R}^{}_\sT {\ovl G}_1^{\perp a}
({\ovl z},{\ovl \xi},\bm{\ovl k}_\sT^2,\bm{\ovl R}_\sT^2,
\bm{\ovl k}^{}_\sT\cdot\bm{\ovl R}^{}_\sT)  \Bigg\} \nn\\[8pt]
=\sum_{a,\ovl a}\; e_a^2 \, \frac{3\,\alpha^2}{2 Q^2} \;z^2 {\ovl z}^2\; 
&A(y) &\frac{1}{M_1 M_2 {\ovl M}_1 {\ovl M}_2} \;G_1^{\perp a}(z,M_h^2) \; 
{\ovl G}_1^{\perp a}({\ovl z},{\ovl M}_h^2) \; \eqcm 
\label{eq:newasym2}
\eea
where
\eqa
G_1^{\perp a}(z,M_h^2) &\equiv &\int d\xi \int_0^{2\pi} d\phi_R^{} \int 
d\bm{k}_\sT^{} \; \bm{k}_\sT^{}\!\cdot\! {\bm{R}}^{}_\sT \; G_1^{\perp a}
(z,\xi,\bm{k}_\sT^2,\bm{R}_\sT^2,\bm{k}^{}_\sT\cdot\bm{R}^{}_\sT) \nn \\
{\ovl G}_1^{\perp a}({\ovl z},{\ovl M}_h^2) &\equiv &\int 
d{\ovl \xi} \int_0^{2\pi} d\phi_{\ovl R}^{} \int d\bm{\ovl k}_\sT^{}
\; \bm{\ovl k}_\sT^{}\!\cdot\! {\bm{\ovl R}}^{}_\sT \; 
{\ovl G}_1^{\perp a} ({\ovl z},{\ovl \xi},\bm{\ovl k}_\sT^2,
\bm{\ovl R}_\sT^2,\bm{\ovl k}^{}_\sT\cdot\bm{\ovl R}^{}_\sT) \; \eqcm
\label{eq:g1perp}
\eea
are weighted moments of the same IFF that appears in the cross section at 
leading twist for two-hadron semi-inclusive DIS off a transversely polarized 
target (see Eq.~(10) of Ref.~\cite{noiDIS}). For simplicity of notation, 
these moments carry no 
further subscripts, as opposed to Eq.~(\ref{eq:h1angle}).

In Eq.~(\ref{eq:newasym2}), the first step follows from first integrating over 
$d\phi^l$, which implies the disappearance of the explicit $\phi_1$ dependence, and 
by performing the $d{\bm q}^{}_\sT$ integration using identities like
\eqa
\int d\bm{q}^{}_\sT \, {\cal F}\left[ \,\bm{\hat h}\!\cdot \!\bm{k}_\sT^{}\,
\bm{\hat h}\!\cdot \!\bm{\ovl k}_\sT^{}\, + \, 
\bm{\hat g}\!\cdot \!\bm{k}_\sT^{}\,\bm{\hat g}\!\cdot \!\bm{\ovl k}_\sT^{} \right]
\; G_1^{\perp a} {\ovl G}_1^{\perp a} &= &\int d\bm{k}_\sT^{} \; k_\sT^i \; 
G_1^{\perp a}(z,\xi,\bm{k}_\sT^2,\bm{R}_\sT^2,\bm{k}^{}_\sT\cdot\bm{R}^{}_\sT) \nn
\\[2pt]
& &\times \int d\bm{\ovl k}_\sT^{} \; {\ovl k}_\sT^i \; 
{\ovl G}_1^{\perp a} ({\ovl z},{\ovl \xi},\bm{\ovl k}_\sT^2,
\bm{\ovl R}_\sT^2,\bm{\ovl k}^{}_\sT\cdot\bm{\ovl R}^{}_\sT) \nn \\[3pt]
\int d\bm{q}^{}_\sT \, {\cal F}\left[ \,\bm{\hat g}\!\cdot \!\bm{k}_\sT^{}\,
\bm{\hat h}\!\cdot \!\bm{\ovl k}_\sT^{}\, - \, 
\bm{\hat h}\!\cdot \!\bm{k}_\sT^{}\,\bm{\hat g}\!\cdot \!\bm{\ovl k}_\sT^{} \right]
\; G_1^{\perp a} {\ovl G}_1^{\perp a} &= &\epsilon_{3ij} \int d\bm{k}_\sT^{} \; 
k_\sT^i \; G_1^{\perp a}(z,\xi,\bm{k}_\sT^2,\bm{R}_\sT^2,
\bm{k}^{}_\sT\cdot\bm{R}^{}_\sT) \nn \\[2pt]
& &\qquad \times \int d\bm{\ovl k}_\sT^{} \; {\ovl k}_\sT^j \; 
{\ovl G}_1^{\perp a} ({\ovl z},{\ovl \xi},\bm{\ovl k}_\sT^2,
\bm{\ovl R}_\sT^2,\bm{\ovl k}^{}_\sT\cdot\bm{\ovl R}^{}_\sT) \; \eqpt
\label{eq:identity}
\eea
The latter integrations can only result in a function of 
$(z,{\ovl z}, \xi, {\ovl \xi}, \bm{R}_\sT^2, \bm{\ovl R}_\sT^2)$
multiplying the products 
$\bm{\hat R}^{}_\sT\!\cdot\!\bm{\hat {\ovl R}}^{}_\sT$ and 
$\bm{\hat R}^{}_\sT\!\times\!\bm{\hat {\ovl R}}^{}_\sT$ of unit vectors, 
respectively, since there are no other available vectors. 

From Eq.~(\ref{eq:newasym2}) it is also easy to check that $G_1^{\perp}$ does
not enter in the integrated, unweighted, cross section. The resulting expression for 
the numerator in Eq.~(\ref{eq:newasym}) becomes
\eqa
\langle\cos(2(\phi_R^{}-\phi_{\ovl R})) \rangle &= &
\int d\xi \int d{\ovl \xi} \int_0^{2\pi} \frac{d\phi^l}{2\pi} \int_0^{2\pi} 
d\phi_R^{} \int_0^{2\pi} d\phi_{\ovl R}^{} \,  
\cos(\phi_R^{} - \phi_{\ovl R}^{}) \nn \\[2pt]
& &\mbox{\hspace{1truecm}} \times \int d\bm{q}_\sT^{} \, 
     \frac{d\sig (e^+e^-\to (h_1h_2)(\bar h_1\bar h_2)X)}
          {dy\,d\phi^l\,dz\,d{\ovl z}\,d\xi\,d{\ovl \xi}\,d{\bm q_\sT^{}}\,
	   dM_h^2\,d\phi^{}_R\, d{\ovl M}_h^2\,d\phi^{}_{\ovl R}} \nn \\[2pt]
&= &\sum_{a,\ovl a}\; e_a^2 \, \frac{3\,\alpha^2}{2 Q^2} \;z^2 {\ovl z}^2\; 
A(y) \; \frac{1}{M_1 M_2 {\ovl M}_1 {\ovl M}_2} \; G_1^{\perp a}(z,M_h^2) \; 
{\ovl G}_1^{\perp a}({\ovl z},{\ovl M}_h^2) \; \eqpt
\label{eq:num2tot}
\eea
The final expression for Eq.~(\ref{eq:newasym}) is
\eqa
A(y,z,{\ovl z},M_h^2, {\ovl M}_h^2) &= & \frac{1}{4} \; \left[ 
\sum_{a,\ovl a}\; e_a^2 \;
\frac{z^2 {\ovl z}^2}{M_1 M_2 {\ovl M}_1 {\ovl M}_2} \; 
G_1^{\perp \,a} (z,M_h^2) \; {\ovl G}_1^{\perp \, a}
({\ovl z},{\ovl M}_h^2) \right] \nn \\[2pt]
& &\times \left[ \sum_{a,\ovl a}\; e_a^2 \, z^2 \, {\ovl z}^2 \, 
D_1^a (z,M_h^2) \; {\ovl D}_1^a({\ovl z},{\ovl M}_h^2) \right]^{-1} \; 
\eqpt
\label{eq:newasym3}
\eea
It is possible to consider the $\bm{q}_\sT^2$ weighting and get $\bm{k}_\sT^2$ 
moments, but we will not do so here. Rather, it is important to remark that the 
weighting factor $\bm{k}_\sT^{}\!\cdot\!{\bm{R}}^{}_\sT$ in Eq.~(\ref{eq:g1perp}) is 
crucial, since the function 
\eq
\int d\xi \int_0^{2\pi} d\phi_R^{} \int d\bm{k}_\sT^{} \; G_1^{\perp a}
(z,\xi,\bm{k}_\sT^2,\bm{R}_\sT^2,\bm{k}^{}_\sT\cdot\bm{R}^{}_\sT)
\ee
does not occur due to parity 
invariance~\cite{Collins:1993kq,Jaffe:1997pv,Bianconi:1999cd}. 
Nevertheless, the chiral-even IFF $G_1^\perp$ can provide a probe of $g_1$ as it 
emerges from the expression of the cross section at leading twist for the two-hadron 
semi-inclusive DIS off a longitudinally polarized target (see Eq.~(31) of 
Ref.~\cite{Bianconi:1999cd}):
\eq
\frac{d\sigma(e \vec p \to e' h_1 h_2 X)_{OL}}
     {d\Omega\,dx\,dz\,d\xi\,d\bm{P}_{h\perp}^{}\, d\bm{R}_T^{}} \propto \Bigg\{ 
     \ldots
{}-\lambda\;|\bm{R}_T|\;A(y)\;\sin(\phi_h-\phi_R)\;
   {\cal F}\left[\,\hat h\!\cdot \!\bm{k}_\sT^{}\,
     \frac{g_{1} \, G_1^{\perp}}{M_1M_2}\right] +\ldots \Bigg\} \; \eqcm
\label{eq:dislong}
\ee
where $\phi_h^{}$ is the azimuthal angle of $\bm{P}_{h\perp}$ (analogously to 
$\phi_1$ in Fig.~\ref{fig:kin}), and $\lambda$ is the target helicity.

This is a good point to make the connection to handedness studies. 
Handedness has been studied for quite some 
time~\cite{Nachtmann:1977ek,Efremov:1992pe,Ryskin:hu} as a means to probe the 
helicity of fragmenting quarks. Clearly, 
$G_1^\perp$ is a similar analyzer of this helicity due to a 
$(\bm{k}_\sT^{}\times
{\bm{R}}^{}_\sT)$ correlation present in the fragmentation process and a 
direct link with the concept of longitudinal jet 
handedness~\cite{Efremov:1992pe} can be made. One can show that the
functions appearing in Eq.~(11) of Ref.~\cite{Efremov:1992pe} are related to
the IFFs dicussed here. For instance, the longitudinal jet handedness is
a linear combination of
functions called $D_1^A$ and $D_2^A$ and is directly proportional to 
$(\bm{k}_\sT^{}\times {\bm{R}}^{}_\sT) \; G_1^\perp (z,\xi,\bm{k}_\sT^2,
\bm{R}_\sT^2,\bm{k}^{}_\sT\cdot\bm{R}^{}_\sT)$.
Similarly, the transversal jet handedness (given by a function called $D_1^T$)
is proportional to $(\bm{k}_\sT^{}\times {\bm{R}}^{}_\sT) \; H_1^\open
(z,\xi,\bm{k}_\sT^2,\bm{R}_\sT^2,\bm{k}^{}_\sT\cdot\bm{R}^{}_\sT)$. 
Although the unintegrated functions $G_1^{\perp}$ and
$H_1^\open$ are directly related to the jet handedness functions of 
Ref.~\cite{Efremov:1992pe}, the asymmetries we have presented here are not easily 
translated to the handedness correlation observables discussed in 
Ref.~\cite{Efremov:1995ff} (different methods of weighting are employed). 
Nevertheless, they should encode similar information
and as such our asymmetry of Eq.~(\ref{eq:newasym3}) could perhaps also 
serve as a measure of a CP-violating effect of the QCD vacuum discussed in 
Ref.~\cite{Efremov:1995ff}. This interesting topic deserves further study. 

The function $G_1^{\perp}(z,M_h^2)$ of Eq.~(\ref{eq:g1perp}) also provides a
probe of the transverse momentum dependent distribution function $g_{1T}$
through asymmetries in the processes $e\, p^\uparrow \to (h_1 h_2) X$
or $p\, p^\uparrow \to (h_1 h_2) X$. However, these are precisely the
processes where also the transversity asymmetries (proportional to $H_1^\open$
and $H_1^\perp$) occur. In fact, the cross section at leading twist for the
two-hadron semi-inclusive DIS on a 
transversely polarized target contains the following terms (see Eq.~(10) of 
Ref.~\cite{noiDIS}):  
\eqa
\frac{d\sigma}{d\Omega\,d x\,d z\,d\xi\,d\bm{P}_{h\perp}\, d M_h^2\,d\phi_R^{}} 
&\propto 
&|\bm{S}_\perp| \, \Bigg\{ \ldots + |\bm{R}_T|\;B(y)\;\sin(\phi_R^{}+\phi_{S_\perp})
  \; {\cal F}\left[\frac{h_1 \, H_1^{\open}}{M_1+M_2}\right]\nn \\[2mm]
&& \mbox{\hspace{-1truecm}} 
 {}+ B(y)\;\sin(\phi_h+\phi_{S_\perp})
   {\cal F}\left[\,{\hat h}\!\cdot \!\bm{k}_\sT^{}\,
     \frac{h_1 \, H_1^{\perp}}{M_1+M_2}\right] \, + \, 
B(y)\;\cos(\phi_h+\phi_{S_\perp})
   {\cal F}\left[\,{\hat g}\!\cdot \!\bm{k}_\sT^{}\,
     \frac{h_1 \, H_1^{\perp}}{M_1+M_2}\right]\nn\\
&& \mbox{\hspace{-1truecm}}  
{}- |\bm{R}_T|\;A(y)\;
   \cos(\phi_h-\phi_{S_\perp})\;\sin(\phi_h-\phi_R^{})\;
   {\cal F}\left[\,{\hat h}\!\cdot \!\bm{k}_\sT^{}\,
                 \,{\hat h}\!\cdot \!\bm{p}_\sT^{}\,
     \frac{g_{1T} \, G_1^{\perp}}{MM_1M_2}\right]\nn\\
&& \mbox{\hspace{-1truecm}}
{}+ |\bm{R}_T|\;A(y)\;
   \sin(\phi_h-\phi_{S_\perp})\;\sin(\phi_h-\phi_R^{})\;
   {\cal F}\left[\,{\hat h}\!\cdot \!\bm{k}_\sT^{}\,
                 \,{\hat g}\!\cdot \!\bm{p}_\sT^{}\,
     \frac{g_{1T} \, G_1^{\perp}}{MM_1M_2}\right]\nn\\
&& \mbox{\hspace{-1truecm}}
{}- |\bm{R}_T|\;A(y)\;
   \cos(\phi_h-\phi_{S_\perp})\;\cos (\phi_h-\phi_R^{})\;
   {\cal F}\left[\,{\hat g}\!\cdot \!\bm{k}_\sT^{}\,
                 \,{\hat h}\!\cdot \!\bm{p}_\sT^{}\,
     \frac{g_{1T} \, G_1^{\perp}}{MM_1M_2}\right]\nn\\
&& \mbox{\hspace{-1truecm}}
{}+ |\bm{R}_T|\;A(y)\;
   \sin(\phi_h-\phi_{S_\perp})\;\cos(\phi_h-\phi_R^{})\;
   {\cal F}\left[\,{\hat g}\!\cdot \!\bm{k}_\sT^{}\,
                 \,{\hat g}\!\cdot \!\bm{p}_\sT^{}\,
     \frac{g_{1T} \, G_1^{\perp}}{MM_1M_2}\right]  \ldots \Bigg\} \eqcm
\label{eq:distrans}
\eea
where $M$ is the target mass with momentum $P^+ = x p^+$. Hence, one 
should carefully project out the azimuthal asymmetry of interest in order to avoid 
contributions from different mechanisms.


\section{\label{sec:end} Conclusions} 

We have studied azimuthal asymmetries in the process 
$e^+ e^- \rightarrow (h_1 h_2) ({\bar h}_1 {\bar h}_2) X$, which function as 
probes for interference fragmentation functions (IFFs). 
The asymmetries arise in
the orientation of the two hadron pairs with respect to each other. We have
presented two asymmetries that are of present-day experimental relevance.
Although the asymmetries probe the correlation of longitudinal and
transverse quark and antiquark spin, respectively, they are to be extracted
from the same experimental data by applying different weights in the form of
trigonometric functions of azimuthal angles. 
The first asymmetry has already been studied by Artru and Collins, but had 
not yet been expressed in terms of the IFF language of 
Refs.~\cite{Bianconi:1999cd,noiDIS,Bacchetta:2002ux}. 
We have also indicated a relation between the function $H_1^\open$,
that occurs in this asymmetry, and transversal jet handedness. 

The second azimuthal asymmetry that we focussed on specifically, has not been
pointed out before and involves the longitudinally polarized quark IFF
$G_1^\perp$, which is related to longitudinal jet handedness.
The asymmetry offers a different way of studying handedness correlations and,
as such, can perhaps be used as a measure of a specific CP-violating effect 
of the QCD vacuum. 
We pointed out that the knowledge of the helicity distribution
function $g_1$ is of help in this respect. 

Extracting IFFs from $e^+e^-$ annihilation will provide for the as yet 
unknown information needed to disentangle the tranversity distribution from 
processes like $e\, p^\uparrow \to (h_1 h_2) X$ or $p\, p^\uparrow \to 
(h_1 h_2) X$. However, we stress that azimuthal asymmetries in these processes
with transversely polarized targets involve combinations like 
$g_{1T} G_1^\perp$, $h_1 H_1^\open$ and $h_1 H_1^\perp$, hence, a careful 
separation of each contribution requires weighting of the cross section with 
the appropriate trigonometric functions.

\begin{acknowledgments}
This work has been supported in part 
by the TMR network HPRN-CT-2000-00130. The research of 
D.B.\ has been made possible by financial support from the Royal Netherlands 
Academy of Arts and Sciences.
\end{acknowledgments}


\end{document}